\documentclass[aps,showpacs]{revtex4}
\usepackage{graphicx}
\newcommand{\qe}{q_e}
\newcommand{\Qop}{\hat{Q} }
\newcommand{\Qdag}{\Qop^{\dag}}
\newcommand{\kndos}{\frac{ (n+ \gamma)  h }{4 \pi \phi_0^2 }  \sqrt{\frac{L}{C} }   }
\begin{document}
\title{Discrete-charge Quantum Circuits and Electrical Resistance}
\author{Constantino A. Utreras D\'{\i}az }
\affiliation{Instituto de F\'{\i}sica, 
Facultad de Ciencias, Universidad Austral de Chile, 
Campus Isla Teja s/n, Casilla 567, Valdivia, Chile}
\email{cutreras@uach.cl}
\date{\today, Valdivia} 
\begin{abstract}
From the theory of quantum $LC$ circuits with discrete charge, and {\em semiclassical} considerations, we obtain approximate energy eigenvalues, depending on the parameter $q_e^2/h$. Next, we include electrical resistance for the quantum $RLC$ circuit, obtaining a relation that strongly reminds us of the Landauer formula.
\end{abstract}
\maketitle
\section{Introduction}

In a series of articles Li and Chen~\cite{LI-CHEN,YOU-LI}  and us~\cite{FLORES,FLORES-UTRERAS,FLORES-UTRERAS2,UTRERAS-FLORES,FLORES-BOLOGNA}, have developed a theory of quantum electrical systems, based on a treating such systems as quantum $LC$ circuits; that is, electrical systems described by two fenomenological parameters: an inductance $L$, and a capacitance $C$. As it is known, when the transport dimension becomes comparable with the charge carrier coherence length, one must take into account not only the quantum mechanical properties of the electron system, but also the discrete nature of electric charge, which leads to the concept of quantum $LC$ circuit with discrete charge.

In this work, we devote section ~\ref{sec-QC} to a brief discussion on some basic facts about quantum circuits with discrete charge, to set up the background for this work. In section ~\ref{sec-Semi} we discuss the {\em semiclassical} approach to the study of quantum circuits, computing the approximate energy eigenvalues using the Bohr-Sommerfeld quantization rules, for two different energy regimes. The approximate energies are expanded in a power series on the dimensionless parameter  $\sqrt{L/C} \,q_e^2/h $. In section ~\ref{sec-RLC} we generalize the equations of motion for the $RLC$ circuit, the resulting equations are equivalent to the damped simnple pendulum. Finally, we find a set of stable solutions, characterized by constant charge and flux, finding a simple relation between the $RLC$ circuit parameters that bears a striking resemblance to the  famous Landauer formula~\cite{IMRY,DATTA}.

\section{Summary of Quantum Circuits}
\label{sec-QC}

In their pioneering article, Li and Chen ~\cite{LI-CHEN,YOU-LI} consider a quantum $LC$ circuit, described by a wavefunction $\Psi(q)$ in the (continuum) charge  ($q$) representation, 
\begin{equation}
\hat H \Psi (q) = -\frac{\hbar^2}{2 L} \frac{d^2}{d q^2} \Psi(q) + V(q) \Psi(q).
\end{equation}
In this description, the charge ($\hat q$ ) and flux ($\hat \Phi$) operators, satisfy the usual commutation rules, $[ \hat q, \hat \Phi ] = i \hbar$, and may be represented by the operators $\hat{ q} =  q $, and $\hat{ \Phi}  =  - i \hbar d/dq $. The charge operator posseses continuum eigenvalues ($q_0$), with delta function eigenstates, just as the single particle quantum mechanical operators $\hat q$ and $\hat p$.

\begin{equation}
\hat{ q} \Psi_{q_0} (q) = q_0 \Psi_{q_0} (q) = q_0 \delta(q-q_0) \nonumber .
\end{equation}


Li and Chen introduce discrete charge eigenstates, whith eigenvalues equal to integer multiples of the quantum of charge (the electron charge, $q_e$), $q_n = n q_e$,
\begin{equation}
\hat{ q} \Psi_{n} (q) = n q_e \Psi_{n} (q) . \nonumber
\end{equation}
In the charge representation, the Schr\"odinger  equation, $\hat H \Psi(q) = E \Psi(q)$, becomes a difference equation, 
\begin{equation}
-\frac{\hbar^2}{2 L q_e^2} \left[ \Psi(q + q_e) + \Psi(q - q_e) - 2 \Psi(q) \right] +  V(q) \Psi(q) = E \Psi(q), \nonumber 
\end{equation}
corresponding to a discrete hamiltonian $ \hat{H} = \hat{H_0} + V(q)$, defined below. It is convenient to define~\cite{LI-CHEN} the discrete charge-shift operators $\hat Q$ and $\hat{ Q}^{\dag}$, and a discrete flux operator $\hat \Phi$,

\begin{eqnarray}
\hat Q ~\Psi(q) &=& \Psi (q + q_e) \nonumber \\
\hat Q^{\dagger} \Psi(q) &=& \Psi(q-q_e) \nonumber \\
\hat{H_0} & = &-\frac{\hbar^2}{2 L \qe^2} ( \Qop + \Qdag - 2) \nonumber \\
\hat \Phi &=& (\hbar/2 i \qe) ( \Qop - \Qdag ).
\end{eqnarray}

It is found that the discrete flux operator $\hat \Phi$ satisfies modified commutation relationships with $\hat q$ and $\hat H_0$ below, due to the discrete nature of electric charge,
\begin{eqnarray*}
\left[ \hat{q}, \hat{\Phi} \right] &=& i \hbar ( 1 + \frac{\qe^2}{\hbar^2} \hat{H_0} ) \\
\left[ \hat{H_0}, \hat{\Phi} \right] &=&  0 \nonumber \\
\left[ \hat{H_0}, \hat{q} \right] &=&  i \hbar \hat{\Phi}. \nonumber
\end{eqnarray*}

In the charge representation, the operator $\hat \Phi$ posesses "plane waves" eigenstates, that is   solutions of the equation $\hat \Phi \Psi(q) = \lambda \Psi(q)$, given by  $\Psi(q) = \exp(i\phi q /\hbar)$, with eigenvalue $\lambda = \sin(q_e \phi/\hbar)/(q_e/\hbar)$, in which $\phi$ is a continuum quantum number, which we call the pseudo flux~\cite{UTRERAS-FLORES}. With this one may define the "pseudo-flux" representation, in which one deals with "wavefunctions" $\Psi(\phi)$, the charge operator becomes a differential operator with respect to $\phi$,  $\hat q = i \hbar \partial /\partial \phi$, and hence the hamiltonian operator becomes a differential operator also, just as in the continuum charge case,

\begin{equation}
\label{Hflux}
\hat H = -\frac{\hbar^2}{2C}\frac{d^2}{d\phi^2}+ \frac{2 \hbar^2}{L q_e^2} \sin^2(q_e \phi/2\hbar).
\end{equation}
Observe that the physical flux is the operator $\hat \Phi = (\hbar/q_e) \sin(q_e \hat \phi /\hbar) \ne \hat \phi $, then, the operator $\hat \phi$ will be called {\em pseudo-flux} from now on. Note also that the physical current operator is obtained from the canonically invariant definition $\hat I = -i [ \hat H, \hat q] /\hbar$, which gives $\hat I = \hat \Phi /L$,  which is  why we consider $\hat \Phi$ to be the physical flux.

\section{Semiclassical study of $LC$ circuit}
\label{sec-Semi}

The electrical engineer has learned to love the simplified description of a system provided by the circuit approximation, when compared with the more complete, but also more complex field description. On the other hand, nature is quantum, we say; however, we describe the behaviour of electrons in modern circuits using the  same basic laws (Kirchhoff laws) as in a classical circuit. There are difficulties on sight for this state of things, since we are now probing nature at very low temperatures, with very pure materials, very tiny currents and strong magnetic fields. Many examples show that things are about to change for the engineer, we have seen flux quantization on superconductors, conductance oscillations, quantum hall effects (integer and fractional), persistent currents and so on~~\cite{IMRY,DATTA}.

It would be very useful to find out to what extent a circuit-like description could be of use for the very small electronic circuits of tomorrow, and what may be retained from such description; a description that has the value of simplicity and familiaruty, which we~\cite{FLORES,FLORES-UTRERAS,FLORES-UTRERAS2,UTRERAS-FLORES,FLORES-BOLOGNA} have been pursuing for some time. Undoubtely, such a descritpion will be able to answer some questionsonly, leaving others unanswered. For example, one area in which the "quantum $LC$ circuit" may give valid results is in the calculation of energy spectra. We have done this for the open electron resonator~\cite{DUNCAN,UTRERAS-FLORES}; but one may question that the calculation was not "simple enough", since it still required to solve the Schr\"odinger equation. 

Now, we propose to go one step further in our simplification, by proposing to use a "semiclassical" approach to the study of quantum circuits. Note that the word {\em semiclassical} will be used in a loose sense, as we shall see. We start from our model hamiltonian in the {\em pseudo-flux} representation (equation~\ref{Hflux} above), and treat all the operators as classical variables; therefore we shall drop the 'hat' from the hamiltonian ($\hat H \to H$) and  the variables ($\hat q  \to q$, $\hat \phi \to \phi$), 

\begin{equation}
\label{Hsemi}
H = \frac{ q^2}{2C} + \frac{2\hbar^2}{L q_e^2} \sin^2(q_e \phi/2\hbar).
\end{equation}

The hamiltonian equations corresponding to eq.~(\ref{Hsemi}) coincide  with the Heisenberg equations for the quantum hamiltonian above, but with all the 'hats' suppressed, i.e.

\begin{eqnarray*}
\frac{\partial H}{\partial q} &=&  \frac{q}{C} = -\dot \phi \\
\frac{\partial H}{\partial \phi} &=& \frac{\hbar}{L q_e} \sin(q_e \phi/\hbar) = \dot q,
\end{eqnarray*}

Observe that $H$ is not a true classical hamiltonian; for example, formally letting $\hbar \to 0$ yields nonsense, unless one takes first the limit $q_e \to 0$, which would be nonsense for us too. Let us define $\phi_0 = \hbar /q_e$, noting that the {\em flux quantum} $\varphi_0 = h/q_e = 2  \pi \phi_0$, so that the hamiltonian and the corresponding equations {\em look classical}, since we eliminated the explicit reference to Plank's constant,
\begin{eqnarray*}
H &=& \frac{ q^2}{2C} + \frac{ 2 \phi_0^2 }{ L } \sin^2(\phi/2 \phi_0 ).\nonumber \\
-\dot \phi &=& \frac{q}{C}  \\
\dot q &=& \frac{\phi_0}{ L } \sin( \phi/\phi_0).
\end{eqnarray*}

The classical equations above ($q$ and $\phi$ are $c$-numbers), may be written as a single second order equation, for the dimensionless  variable $\theta = \phi/\phi_0$, which coincides with the equation for the {\em simple pendulum}, with frequency $\Omega_0 = 1/\sqrt{LC}$, $\ddot \theta = -\Omega_0^2 \sin(\theta)$, which has been extensively studied form many years, becoming a standard textbook exmpla in classical mechanics~\cite{LANDAU}. We are interested in this {\em classical} equation, as a means to obtain the eigenstates of the quantum hamiltonian eq.~\ref{Hflux}; the method is the Bohr-Sommerfeld quantization rules (WKB approximation) of the old quantum theory, i.e., $\oint q d\phi = (n +\gamma) h$ ($\gamma$ is a constant, we take $\gamma = 1/2$). 

Observe that the constant $\phi_0$ defines a scale of energy for our quantum $LC$ system, namely $\phi_0^2/L$, therefore,, there exist two energy regimes, the {\em low energy regime}, defined by $E< 2\phi_0^2L$, and the {\em high energy} regime, defined by $E> 2\phi_0^2/L$.

\subsection{Low energy states: $E< 2\phi_0^2L$  }

We consider here the calculation of the energy spectrum of the quantum $LC$ system, in the 

\begin{eqnarray*}
q &=& C \dot \phi = C \phi_0 \dot \theta,  \\
E &=& C \phi_0^2  \left( \frac{\dot \theta^2}{2} + 2\Omega_0^2 \sin^2(\theta/2 ) \right).
\end{eqnarray*}
Let $\theta = \theta_0$ be the maximum amplitude, defined by $  \sin^2(\theta_0/2) = E / ( 2 C \phi_0^2 \Omega_0^2 )   = E/(2\phi_0^2/L)$, then

\begin{equation}
\dot \theta = \pm 2 \Omega_0 \sqrt{ \sin^2(\theta_0/2) -  \sin^2(\theta/2) }. \nonumber
\end{equation}
To find the (quantum) energy eigenstates we need the phase-space integral $I$, 
\begin{equation}
I = \oint q d\phi = C \phi_0^2  \oint  \dot \theta \, d\theta = 2 \sqrt{\frac{C}{L} } \phi_0^2 \oint \sqrt{ \sin^2(\theta_0/2)-\sin^2(\theta/2)   } \, d\theta \nonumber
\end{equation}
which may be expressed as an elliptic integral of the second kind, expressed in the standard form by the transformation $\sin (\xi) = \sin(\theta/2)/\sin(\theta_0/2)$, defining further $k = \sin(\theta_0/2)$, we get

\begin{eqnarray*}
I &=& 16 \sqrt{\frac{C}{L} } \phi_0^2 k^2 \int_0^{\pi/2} \frac{ \cos^2(\xi)~d\xi }{\sqrt{1 - k^2  \sin^2(\xi)} }.
\end{eqnarray*}
This integral is given in the tables ~\cite{GRADSHTEYN}, page 162 (2.584, \#6), in terms of the complete elliptic integrals, ${\bf K}(k) = {\bf F}(\pi/2,k)$ and $ {\bf E }(k) = {\bf E}(\pi/2,k)$,

\begin{equation}
I =  16  \sqrt{\frac{C}{L} } \phi_0^2  \left( {\bf E}(k ) - (1 - k^2) {\bf K}(k )\right).
\end{equation}
To obtain approximate expressions for the energy eigenvalues, we use the series expansion for ${\bf K}(k)$ and ${\bf E}(k)$ (Gradshteyn~\cite{GRADSHTEYN}, pp. 905), valid for $k << 1$, we get 

$$
I \approx  4 \pi \sqrt{ \frac{C}{L} } \phi_0^2 k^2 \left( 1  + \frac{k^2}{8}  \right).
$$

Impose now  the quantization rules of the old quantum theory, $I = (n + \gamma) h$, to obtain the energy from 
\begin{equation}
I \approx  4 \pi \sqrt{ \frac{C}{L} } \phi_0^2 k^2 \left( 1  + \frac{k^2}{8}  \right) = (n + \gamma) h.
\end{equation}

Assuming $k << 1$ we have, as a first approximation,

$$
(k_n^0)^2 =  \kndos ,
$$
which gives the usual result for the energy, $E_n^{0} = (n + \gamma) \hbar \Omega_0$. Now compute the corrected
$k_n$, 
$$
k_n^2 = \kndos \left( 1 - \frac{1}{8}\kndos~ \right) .
$$

The energy levels become (using $\gamma = 1/2$, as indicated previously), up to second order 

\begin{eqnarray}
E_n  &\approx& \hbar \Omega_0 (n + \gamma) \left( 1 - \frac{1}{8}\kndos \right) \nonumber \\
& =& \hbar \Omega_0 (n + \gamma)
\left[ 1 - \frac{\pi (n + \gamma) }{8} \frac{q_e^2}{h} \sqrt{ \frac{L}{C} } ~ \right].
\end{eqnarray}

This result becomes $E_n = (n + 1/2) \hbar \Omega_0$, in the continuum charge approximation $q_e \to 0$. Notice also that, on the next approcimation, the result depends on the ratio of the Landauer conductance $q_e^2/h $, and $LC$ conductance $\sqrt{C/L}$, an interesting  result, as we shall see later on.

\subsection{High energy states: $E> 2\phi_0^2/L$, or $C$-design case}

The so-called $C$-design case corresponds to the 'large capacity' case, in which the electrostatic part
dominates. We have seen that the 'velocity' $\dot \phi$ may be written as

$$
\dot \phi = \pm \sqrt{\frac{2 E}{C}} \left[ 1 - \frac{ 2 \phi_0^2}{ L E} \sin^2(\frac{\phi}{2 \phi_0}) \right]^{1/2}.
$$

Consider the case in which the energy $E >> \phi_0^2/2L$, i.e., in which the electrostatic energy dominates, the so-called $C$-design. It is convenient to define the parameter $\lambda = (2\phi_0^2/L)/E$, to write the expression for the electric charge $q$, 

\begin{equation}
q =  \pm \sqrt{2 C E} \left[ 1 - \lambda \sin^2( \frac{\phi}{2 \phi_0}) \right]^{1/2}.
\end{equation}
Observe that in the present case, the  pseudo flux variable $\phi$ grows without bounds, but the system is periodic with period $\Delta \phi = 2\pi \phi_0$. To compute the action integral $I$, it is convenient to change variables to $\alpha = \phi/2 \phi_0$,

\begin{equation}
I = \oint q ~d\phi =  \sqrt{2 C E} \, 2 \phi_0  \int_0^{\pi}  \left[ 1 - \lambda \sin^2(\alpha) \right]^{1/2} \, d\alpha = 2 \phi_0 \sqrt{2 C E} \,   {\bf E} (\pi, \sqrt{\lambda} ) , \nonumber
\end{equation}
in which we notice that and that ${\bf E}(\pi,\sqrt{\lambda})$ is an elliptic integral~(\cite{GRADSHTEYN} equation 8.121.4, pag. 906), that ${\bf E}(\pi,\sqrt{\lambda}) = 2 {\bf E}(\pi/2,\sqrt{\lambda}) = 2 {\bf E}(\sqrt{\lambda})$, and also $2 \phi_0 \sqrt{ 2 C E} =  ( 4\phi_0^2/ \sqrt{ \lambda } ) \sqrt{C/L }$, hence, we have

\begin{equation}
I = \frac{8 \phi_0^2}{\sqrt{\lambda}} \sqrt{ \frac{C}{L}} {\bf E} (\sqrt{\lambda}). \nonumber
\end{equation}
Now, use the series expansion of ${\bf E}(\sqrt{\lambda})$~(\cite{GRADSHTEYN}, equation 8.113, pag. 905), and replace $\phi_0 = h/2\pi q_e$,

\begin{equation}
I = \frac{1}{\pi \sqrt{\lambda}} \sqrt{ \frac{C}{L} } \frac{h^2}{q_e^2}  \left[  1 - \frac{\lambda}{4} -\frac{3 \lambda^2}{64 } 
- \frac{5 \lambda^3}{256} + \cdots \right] . \nonumber
\end{equation}
Now, impose $I = n h$ (use $\gamma = 0$ for the Maslov index), and compute the first approximation $\lambda_0$,

\begin{equation}
\frac{1}{\lambda_0} = \frac{L}{C} \frac{n^2 q_e^4}{2 \hbar^2}  = \left( \frac{L}{2 \phi_0^2}  \right)  \frac{(n q_e)^2 }{2 C} 
\end{equation}

The condition $\lambda < 1$ becomes $n > \sqrt{C/L} /( \pi q_e^2/h ) $ , for the quantum number $n$. This is interesting, since $\sqrt{C/L}$ is the conductance associated to the $LC$ circuit, and $ q_e^2/h $ is the conductance quantum, appearing in the Landauer formula, this ratio appears as an expansion parameter. Define the dimensionless ratio $\mu = \lambda /\lambda_0 $, then the following relation is equivalent to the quantization condition $I = n h$,

\begin{equation}
\sqrt{ \mu} = \left[ 1 - \frac{\lambda_0 \mu }{2^2} - 3 \frac{\lambda_0^2 \mu^2}{8^2} - 5 \frac{\lambda_0^3 \mu^3}{16^2} + \cdots \right]. \nonumber
\end{equation}
Now, assume the expansion $\mu = 1 + a \lambda_0 + b \lambda_0^2 + c \lambda_0^3 + \cdots$, and insert into the equation above (after squaring both sides); the result for $\lambda$ becomes
\begin{eqnarray*}
\lambda &=& \lambda_0 \left( 1 - 2 \lambda_0 + 29 \lambda_0^2/32 + \cdots \right) , \\
\lambda^{-1} &=& \lambda_0^{-1} \left( 1 + 2 \lambda_0 + 99 \lambda_0^2/32 + \cdots \right)
\end{eqnarray*}
then the approximate energy eigenstates are

\begin{eqnarray}
E_n &=& \frac{2 \phi_0^2}{L \lambda} = \frac{2 \phi_0^2}{L \lambda_0} + \frac{4 \phi_0^2}{L} + \frac{99 \phi_0^2 \lambda_0}{16 L} \\
&=& \frac{(n q_e)^2}{2 C}  +  \frac{4 \phi_0^2}{L}  + \frac{99}{32} \left( \frac{ \phi_0^2}{L} \right)^2  \frac{C}{(n q_e)^2}
\end{eqnarray}

\section{Electrical resistance and circuit laws}
\label{sec-RLC}

The problem of electrical resistance  at the mesoscopic level is a very subtle one. The problem has been studied by Landauer and others~\cite{IMRY}, who obtained a relationship between conductance and the transmission coefficient in a one-dimensional scattering experiment. The role of the contact resistance, as well as that of the different physical regimes in which one may observe a system, complicate the description of a system.

\subsection{The generalization}

Here we propose to study a generalization of our equations to include electrical resistance, in a manner similar to what happens in a classical circuit. To carry out this generalization, we may proceed by replacing the physical  current $I$ occurring in a {\em mesoscopic circuit} by the value ${\cal I}$, and adding a 'resistance' term, in the same way as it is done in a classical circuit, i.e. by adding a {\em potential drop} $\Delta {\cal V} =R {\cal I} = R\phi/L$; then the equations become

\begin{eqnarray}
\label{eqRes}
-\dot \phi - R {\cal I} &=& \frac{q}{C} \\
\dot q & =& \frac{\phi_0}{L} \sin( \phi/\phi_0). \nonumber
\end{eqnarray}

In this way, we have a generalization that preserves  the form of Faraday's law of induction, $\varepsilon = -d\phi/dt$, and Ohm's law, $\Delta {\cal V} = R{\cal I}$; which is rather nice for engineering applications. Now, we must decide which value should we assign to the ${\cal I}$ appearing on equation~\ref{eqRes}. We note that there are two possible choices for it, namely, we may identify

\begin{itemize}
\item ${\cal I} = \dot q$, or
\item ${\cal I} = \phi/L$.
\end{itemize}

Let us consider first the (possible) identification ${\cal I} = \dot q$. In this case, the system is described by

\begin{eqnarray}
\label{eqRes1}
-\dot \phi - R \frac{\phi_0}{ L} \sin( \phi/\phi_0)  &=& \frac{q}{C} \\
\dot q & =& \frac{\phi_0}{L} \sin( \phi/\phi_0). \nonumber
\end{eqnarray}

If we study the equilibrium solutions of eq.~\ref{eqRes1}, we see that $\dot q =0 $ implies $\phi = n \pi \phi_0/2 $, and that $\dot \phi = 0$, implies that the charge $q = 0$. 
\par
Consider now the second choice, our proposal, in which ${\cal I}= \phi/L$. The quantity ${\cal I}$ may be called a {\em pseudo-current}, since $\phi$ is already called {\em pseudo-flux} by us; notice then that  ${\cal I} \ne \dot q$, which is ok, since the true current is $I= \dot q$. Our new circuit  equations become

\begin{eqnarray}
\label{eqRes2}
-\dot \phi - R \frac{\phi}{L}   &=& \frac{q}{C} \\
\dot q & =& \frac{\phi_0}{L} \sin(\phi/\phi_0). \nonumber
\end{eqnarray}


To make our generalization more compelling, we prove that it may be obtained from a lagrangian formulation. To see this, we write the equation of motion as a single, second order equation,

\begin{equation}
L \ddot \phi + R \dot \phi + \frac{\phi_0}{C} \sin( \phi/\phi_0) = 0.
\end{equation}

This equation may be obtained from the Lagrangian description, with the use of a "dissipation function", as  described by Landau~\cite{LANDAU}. Let ${\cal L}$ be the lagrangian, and ${\cal F}$ be the dissipation function, then
\begin{eqnarray}
{\cal L} &=& \frac{C \dot \phi^2}{2} - \frac{2 \phi_0^2}{L} \sin^2(\phi/2\phi_0) \\
{\cal F} & =& \frac{R \dot \phi^2}{2} \\
\frac{d}{dt} ( \frac{\partial {\cal L}}{\partial \dot \phi}) - \frac{\partial {\cal L} }{\partial \phi} 
&=& -\frac{\partial {\cal F}}{\partial \dot \phi} 
\end{eqnarray}
The last equation then is exactly our proposed equation. This is a good generalization, in the sense that the energy $E$, defines in the usual way, 

\begin{equation}
E = \frac{C \dot \phi^2}{2 } + \frac{2 \phi_0^2}{L} \sin^2(\phi/2\phi_0),
\end{equation}
decreases, or stays constant (if $\dot \phi = 0$), since
\begin{equation}
\frac{dE}{dt}  = - R \dot \phi^2.
\end{equation}

The new equations have also the correct classical limits, obtained for $q_e \to 0$ and small pseudo-flux ( $\phi << \phi_0$), they also  have more interesting structure; for example, the equations posess nontrivial solutions with constant pseudo flux and constant charge, to be discussed below.

\section{Relation to Landauer formula}

As it was indicated above, the system of equations ~\ref{eqRes2} posseses constant flux and charge solutions, which we find by assuming $\dot q = 0$ and $\dot \phi = 0$, we obtain $\phi_n /\phi_0 = n \pi $, for integer $n$,  and we get the condition $-R\phi_n/L = q_n/C$, i.e., $q_n = - RC \phi_n/L = -n \pi  RC\phi_0/L $. The solutions having  even $n$ are stable, and those with odd $n$ are unstable, as it is shown by numerical calculations: the solutions for odd $n$ "decay" onto the $n-1$, i.e., dissipating energy in the process, while $\dot \phi \ne 0$. Hence, from now on we keep only the stable, even $n$ solutions (replacing $n$ by $2n$). We have shown then, that there exist {\em stable equilibrium solutions} of the equations of motion~\ref{eqRes2}, which may be expressed as
\begin{eqnarray*}
\phi &=& 2 n \pi \phi_0  \\
q & = - & 2 n q^* \\
q^* &=& \frac{\pi \phi_0 R C}{ L} .
\end{eqnarray*}

These solutions show that the {\em pseudo} flux $\phi$ is an integer multiple $ 2 \pi \phi_0$, and that the charge is an even integer multiple of the charge $q^* =  \phi_0 RC/L$. The physical, equilibrium charge in the system is $q$; therefore, this charge $q$ must be an integer multiple of the unit of electronic charge ($q_e$), i.e., it should be written as $q = m (-q_e)$, in which  $m$ is another integer number, not necessarily equal to $n$, as we shall see below. We have
$$
-m q_e = -\frac{2 n RC \pi \phi_0}{L},
$$
then we obtain
$$
\frac{RC}{L} = \frac{ m}{2n} \frac{q_e}{\pi \phi_0 } = \frac{m}{2n} \frac{2 q_e^2}{h} = \frac{ m}{n} G_{Landauer}.
$$

The quantity $G_{Landauer} = q_e^2/h = 3.875\times 10^{-5}~ \Omega^{-1} $ is the {\em Landauer conductance}. This is a remarkable result, obtained by a simple generalization of the  equations of motion to account for electrical resistante. Notice also that the quantity $C/L$ is actually the product of the $C-$ conductance and an $L-$ conductance, in other words, recalling that the capacitive and inductive conductances at frequency $\Omega_0$, $G_C = \Omega_0 C$, and $G_L = 1/\Omega_0 L$ (at frequency $\omega$) satisfy $G_C G_L = C/L$, we may write, or that $G_{LC } = \sqrt{C/L}$ may be viewed as the conductance of thew $LC$ circuit.

\begin{equation}
\label{eqGeff}
G_{eff} = R G_{LC}^2  =   \frac{m}{n} \frac{q_e^2}{h } =  \frac{m}{n} \, G_{Landauer}.
\end{equation}

We may say that the circuit should show an {\em effective conductance} $G_{eff}$, which is a rational multiple of the Landauer conductance $G_L$. We have established a relationship between the parameters of a mesoscopic system, that bears a strong resemblance to the Landauer formula, this result comes about because we have insisted on charge quantization. This is analogous to what happens in the old quantum theory, in which the angular momentum quantization, applied to the Bohr atom actually {\em selects} the energies. In our case, our conditions selects the cases in which we have both flux and charge quantization, resulting in equation~\ref{eqGeff}.


The previous relation may be generalized slightly if one considers the influence of an external, constant electromotive source $\varepsilon_0$, in this case, the circuit equations become

\begin{eqnarray}
\label{eqRes3}
\varepsilon_0   &=& \frac{q}{C} + \dot \phi + R \frac{\phi}{L}  \\
\dot q & =& \frac{\phi_0}{L} \sin(\phi/\phi_0). \nonumber
\end{eqnarray}
Now, let us impose the condition $\dot q = 0$, and $\dot \phi = 0$, then $\phi = n \pi \phi_0$, as before, and the charge becomes
\begin{equation}
q = C \left[ \varepsilon_0 - n \pi \frac{R \phi_0 }{L} \right].
\end{equation}
Now, as in the previous paragraphs, we consider only the stable (even n) solutions, i.e., replace $n \to 2n$, and impose the condition $q = -m q_e$, for integer $m$, then we obtain the relation
\begin{equation}
m G_{Landauer} = - C \varphi_0 \varepsilon_0 + n G_{eff}
\end{equation}

Finally, if one considers the system as driven by an external, time-dependent perturbation, $v(t)$, the equations of motion may now be written by simply adding the external source, as below
\begin{eqnarray}
\label{eqRLC}
\dot \phi &=& - \frac{R \phi}{L} - \frac{q}{C} + v(t) \\
\dot q & =& \frac{\phi_0}{L} \sin(\phi/\phi_0). \nonumber
\end{eqnarray}
It it clearly seen it is not possible to obtain an equation fot the charge variable $q$ alone, but that it may easily done for the pseudo flux variable $\phi$, the result has a  simple and well known form, namely, that of the forced simple pendulum with dissipation. These equations describe may be used to describe many different physical systems, including mechanical, electrical, superconductors, and so on, and it will be considered elsewhere;for the variables of our {\em semiclassical} $RLC$ circuit the equations are, which we write for the sake of completeness

\begin{equation}
\ddot \phi + \frac{R \dot \phi}{L} + \frac{\phi_0}{L C} \sin(\phi/\phi_0) = \dot v(t).
\end{equation}
\section{Final Remarks}
We started from a description of the quantum $LC$ circuit, and introduced the {\em semi classical} approximation to compute the energy eigenvalues of the system. We computed the energy eigenvalues for two different energy regimes. First, we considered the low energy regime $E< \phi_0^2/L$, and next, the high energy regime $E>\phi_0^2/L$. In both cases we computed the energy as a power series in a small dimensionless parameter, which may be defined as $\sqrt{L/C} \,( q_e^2/h)$, i.e., the Landauer conductance times the impedance $\sqrt{L/C}$ of the $LC$ circuit. In this way, we have shown in a very simple way the important role of $q_e^2/h$, and its relation to charge quantization on the $LC$ circuit.

Next, we generalized the equations of motion of the $LC$ circuit to account for electrical resistance, which is equivalent to generalizing the usual Kirchoff circuit laws, to account for charge discretization and electrical resistance; the equations obtained are equivalent to the simple pendulum with forcing. The $\sin (\phi/\phi_0)$ term  in the equations of motion  generates a set of stable constant flux and charge solutions for the unforced mesoscopic $RLC$ circuit, which, when charge quantization is imposed on the solutions, generates a condition to be satisfied by the parameters of the system~\ref{eqGeff}. This equation bears a striking resemblance to the Landauer formula and also to the quantization of conductance of the quantum  Hall effect. This is a most amazing result, since it has been obtained from a generalization of the circuit equations, and simple considerations of charge  quantization.

\section{Acknowledgements}

The author aknowledges the finantial support provided by FONDECYT Grant \# 1040311, and DIDUACH Grant \# S-2004-43. Use has been made of the software package Maxima (http://www.sourceforge.net). Thanks are also due to Profs. A. Zerwekh (Universidad Austral de Chile) and J. C. Flores (Universidad de Tarapac\'a), for discussions and comments on this work.

\end{document}